# A Visual Grammar Approach for TV Program Identification

**Tarek ZLITNI[1] and Walid MAHDI[2]**

[1]Mir@cl, Multimedia, Information Systems and Advanced
Computing Laboratory Higher Institute of Computer Science and Multimedia
Sfax BP 3021, 69042 TUNISIA, University of Sfax
*tarek.zlitni@isimsf.rnu.tn*

[2] Mir@cl, Multimedia, Information Systems and Advanced
Computing Laboratory Higher Institute of Computer Science and Multimedia
Sfax BP 3021, 69042 TUNISIA, University of Sfax
*walid.mahdi@isimsf.rnu.tn*

**Abstract:** *Automatic identification of TV programs within TV streams is an important task for archive exploitation. This paper proposes a new spatial-temporal approach to identify programs in TV streams in two main steps: First, a reference catalogue for video grammars of visual jingles is constructed. We exploit visual grammars characterizing instances of the same program type in order to identify the various program types in the TV stream. The role of video grammar is to represent the visual invariants for each visual jingle using a set of descriptors appropriate for each TV program. Secondly, programs in TV streams are identified by examining the similarity of the video signal to the visual grammars in the catalogue. The main idea of identification process consists in comparing the visual similarity of the video signal signature in TV stream to the catalogue elements. After presenting the proposed approach, the paper overviews the encouraging experimental results on several streams extracted from different channels and composed of several programs.*

**Keywords:** Visual grammar, TV programs identification, TV stream segmentation, video signature.

## 1. Introduction

Face to the ubiquity of large volumes of digital videos, effective and rapid access to multimedia information has become a difficult endeavour. This difficulty created a growing demand for efficient tools that offer direct access to unstructured video contents, which in turn require automatic video segmenting and indexing methods. Indeed, several approaches were proposed to segment/index particular video types, cf. [1] [4] for sports videos, [5] for news videos. However, being dependent on the video type (e.g., documentaries, sports, films, news, etc), the effectiveness of the proposed approaches cannot be sustained when the video stream is composed of several programs of different types. Hence, to profit from the robustness of these approaches, one must first segment the video stream by identifying the types of video programs it contains. In fact, besides video stream segmentation for efficient indexing of its programs, video type identification can be exploited in a wide range of applications like controlling the respect of broadcasting policies and agreements by TV channels or video copies detection. For instance, companies invest several millions to reserve particular time spaces for their advertisements; in this context, an automatic content/type identification tool can assist these companies in verifying if their advertisements are broadcasted during the agreed periods.

The above mentioned applications and importance as a pre-treatment step for efficient indexing, we present in this paper a new approach for video TV stream content/type structuring. Our approach constructs first a reference catalogue composed of visual markers (visual jingles) for various TV programs associated to their video grammars. Each grammar defines a set of visual descriptors appropriate to a particular TV program. In the second stage, our approach exploits this catalogue to identify programs in TV streams based on the visual similarity between the visual grammars and the given video stream signal.

The rest of the paper is structured as follows: in Section 2, we introduce the context and motivation of our approach. In Section 3, we describe the two-step proposed approach. We present the results of a preliminary experimental evaluation in Section 4. We conclude the paper with discussions of the proposed solution and an outline of future work.

## 2. Context and Motivation

Video segmentation is an essential step in any video exploitation application. On the other hand, in spite of their satisfying results, the performance of the majority of current video segmentation approaches depends on the video type (news program, action film, documentary film, sport program, etc). This dependency on the video type is explained essentially by the fact that currently proposed approaches are based on the production rules of the video. For example, the work in [5] segments through face detection because it considers that the occurrence of a person shot indicates mostly a subject change in news programs; Mahdi et *Al*. [11] relies on the detection of cinematic rhythm to segment fiction films; the works presented in [4] and [1] are based on prior knowledge about production rules specific to sports programs in order to detect particular events e.g., attacks, fowls, etc.

Overall, prior knowledge of the video type is required for a good semantic exploitation of a video segmentation. However, this knowledge is often unavailable in particular in archive videos, videos widely distributed over the web, etc. Furthermore, videos are generally integrated into heterogeneous and long streams, like for instance those incorporated into channel broadcasts. In this latter case, a video sequence can include a set of programs of various types (documentaries, news programs, films, etc.) and/or inter-programs (spotlights, commercials, etc.) without any indication about the boundaries between them.

Given this context, two levels of segmentation are required for an efficient video structuring (Figure 1). The first level, called inter-segmentation, is dedicated to identifying a program or an inter-program in the TV stream. The second level, called intra-segmentation, aims to segment the identified program into a set of temporal units and semantic entities.

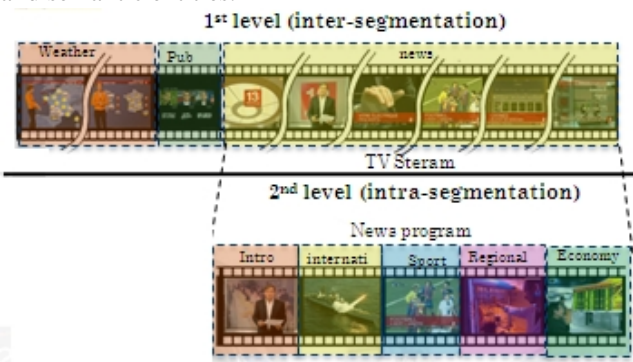

**Figure 1.** A two-level structuring process for TV streams.

Besides efficient intra-segmentation, inter-segmentation for TV streams could be the object of various applications. In fact, the identification of TV programs could serve to develop control mechanisms for TV channels broadcasting. On this matter, the audiovisual authorities (such as the superior audiovisual council (*CSA*) in France) impose a set of policies for TV broadcasting and control the respect of these ones by TV channels. As an example, the *CSA* [3] imposes that total duration of advertisements within one hour of broadcasting must not exceed 4 minutes for the public channels and 6 minutes for the private ones. Another control example is in the context of respect of political pluralism, where the candidates must have the same duration in television electoral campaigns.

Besides the authorities of audiovisual control, companies paying millions to reserve particular time spaces with high audience for their advertisements need a means to control whether the agreement was respected. In this context, such companies would find it more advantageous to use automatic tools than a manual verification to control tens of channels with a big risk of inattention of 15 seconds from a human agent, which can have devastating consequences.

In another context, channels diffusing 24 hours a day would have digital archives with thousands of gigabytes of data. However, even though most channels diffuse around the hour (24/24), programs are diffused at least twice a day. Consequently, identifying the repeated contents (replay programs) as well as the inter-programs and their removal during the digital storage phase reduce a large part of these archives.

## 3. Video type identification approach

A great deal of multimedia content identification research focuses on identifying video copies and web databases [10] [17], whereas a few works has been interested in programs identification in long TV streams.

Among the latter works, we refer to the work presented in [2] which proposes a method to detect the repeated sequences in TV streams. This method is based on inter-program detection (commercials, jingles, spots …) to split the TV stream and extract useful programs. It detects repeated sequences using a clustering technique that groups similar audio/visual features vectors of key-frames.

On the other hand, Naturel presents in [12] a two-step method to structure TV streams. The first step creates a hash-table that contains the key-frames signatures. Then, the second step uses signature similarity comparison between the stream signal and the hash-table to identify the various programs.

Contrary to these two methods, we propose a video grammar-based method that relies on TV channel "graphic charter" to generate an appropriate grammar for visual jingles. The produced grammar provides for the detection of the generics (*i.e.,* the starting points) of programs in a TV stream.

As illustrated in Figure 2, our approach proceeds mainly in two steps. The first step constructs a reference catalogue containing visual grammar definitions. Each grammar is represented in a spatial-temporal signature format. The second step identifies the program types in TV streams based on grammar components similarity measurement.

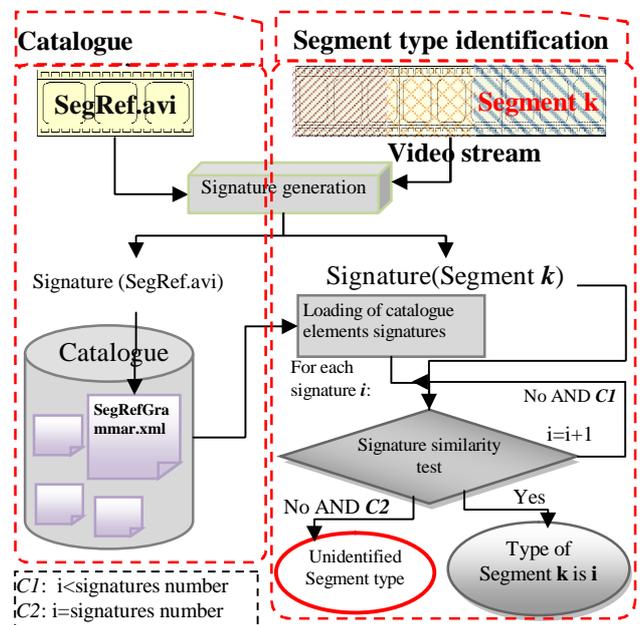

**Figure 2.** Our TV program type identification approach.

### 3.1 Construction of the video reference catalogue

The reference catalogue contains a set of visual markers associated with their grammars. The markers are the visual jingles whose principal role is to identify visually the starting moment of a given program, as perceived by the viewers.

Generally, the graphic chart of the TV channels uses jingles for a particular program during a long period.

Consequently, based on the fact that the instances of a given program have the same visual marker, we proceed to represent the visual characteristics of an instance of a particular type to identify thereafter the other ones.

### 3.1.1 State of art of visual content representation

Several researches have dealt with visual content signature generation. A signature is a compact description that constitutes the start point for the similarity detection of the visual contents. Signature determination allows a direct indexing that facilitates the identification of similar contents. Generally, a signature generation requires mainly two steps: First, detecting the significant low-level image features, and secondly characterizing descriptors in compact format.

According to the most realized researches in this context, we cite two main types of relevant image primitives used to compute the signature: point of interest [15] [18] and color [7] [8]. Works based on first class of primitives (*POI*) were designed using different detectors such as Harris [6] or SIFT detector [15]. These methods are based on a common ground: detection and description of the most relevant points in the image, which gives more information than others [15]. As an example, we cite the work of Stoettinger [14] which uses a color version of *POI* detector for image retrieval.

For the second class relying on the color feature, the main idea is to represent the image color information in a distinctive and reduced format. For example, [7] [8] use the ordinal measure method to represent the intensity distribution within the image. These methods proceed first by computing the intensity average of $N$ blocks in the given image. Secondly, the set of average intensities is sorted in ascending order and the rank is assigned to each block referring to their mean intensity value. The ordinal measure is expressed by the ranked sequence. On the other hand, another type of methods exploiting the color information to create the signature use color coherent vectors *CCV* [16] and Vector quantization [9].

In summary, most existing video signature computing employ feature vector extracted from a single frame for two reasons: some approaches are proposed to CBIR systems, and others rely on key-frame signature generation to detect web videos copies. Hence, the majority of these approaches do not consider the temporal aspect even though it improves the signature efficiency.

### 3.1.2 Audiovisual grammar and video signature generation

The concept of grammar is defined as a set of formalisms allowing representing the relations which can exist between a set of entities. This formalization makes it possible to represent data in structured and significant way for a better semantic interpretation. Inspired from this definition, in the audiovisual field, the grammar notion is a recent concept whose aim is to define an appropriate style to the TV channels and to deduce the typical structures of TV programs. Hence, a video grammar could be exploited in a multitude of multimedia applications. As examples of these applications, we cite the identification of TV programs type [19], the characterization of a particular event in a sport program (substitution, goals …). Indeed, video grammar is defined by the set of the visual and sound entities and their relations which carry out to characterize a visual identity for a particular TV channel. This identity is created by the conception of a suitable graphic style. Graphic components of this style are generally recurring for a long duration (minimum a year) and their application follows a logic specified by the "graphic chart" of TV channel. Consequently, the grammar generation consists first of detecting the visual invariants, then describing them in a formal way and finally deduce the semantic interpretation for their appearance. For example, in the case of sport video grammar, the detection of cautioned player event requires the extraction of the markers zones of text as well as the visual invariants to symbolize the yellow card.

### 3.1.3 A spatial-temporal video signature

Our video signature generation method is based on visual invariants (forms, colors …) (Figure 3). This approach relies on the fact that the TV channels programs (such as news or sport programs) use distinctive graphical components to identify them visually. As a result, we exploit visual grammars characterizing instances of the same program type in order to identify the various program types in the TV stream. The main role of this grammar is to represent the visual invariants for each visual marker (jingle) using a set of descriptors appropriate for each TV program.

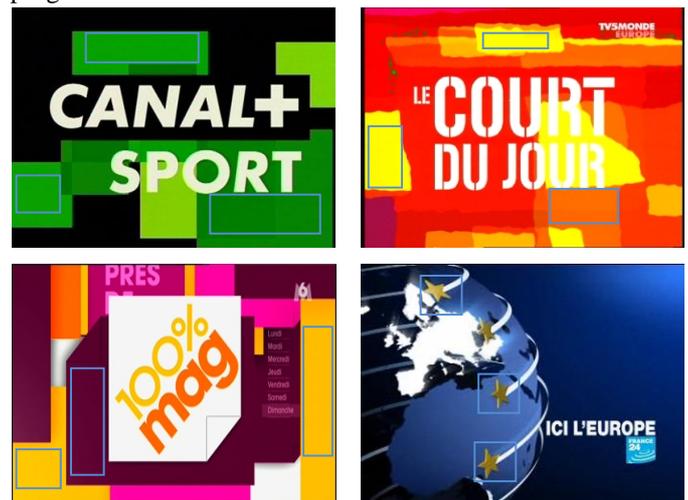

**Figure 3.** Samples of visual invariants of different TV programs.

For an efficient representing of the visual invariants, it is necessary to extract the most relevant features that can characterize them. In fact, the signature generation process creates a compact description of these features while respecting two crucial properties: robustness and uniqueness. These properties guarantee the discriminative effect for distinguishing video content and ensure the capability of noise tolerance. For the robustness property, a signature must not vary when the video sequence contains for example an insignificant signal noise or slight luminosity variation. For the uniqueness property, two different video contents must have two different signatures. In this sense, every semantically different video segment should possess a unique signature. Consequently, we propose a new spatial-temporal method to generate a signature for video segment identification that preserves these two properties.

Indeed, contrary to most proposed approaches, we generate video signature from a set of frames of the audiovisual segment and not a single (key) frame. This fact may influence on the signature efficacy (*i.e.,* the uniqueness property) as two videos with similar key-frames do not necessarily have the same or similar content.

In order to overcome this deficiency, we opt for a bi-dimensional signature. The main idea is to create a spatial-temporal signature (1) where the generation process is carried out from a set of frames, separated by a definite time step $T_{step}$. That is, the generation process provides different levels of signature discrimination. Two signatures will be similar/ dissimilar on three levels: $N_{frame}$, $T_{step}$ and $SigF$ which can further guarantee the uniqueness property.

$$Vsig = \left( \left| \overrightarrow{SigF} \right|, N_{frame}, T_{step} \right) \qquad (1)$$

Our signature generation process starts by select a number of frames $N_{frame}$ and a temporal step ($T_{step}$) which separates them. Secondly, for each frame of the $N_{frame}$ selected frames, we compute $\overrightarrow{SigF}$ the low-level characteristics vector derived from this frame. The signature of a video segment is defined by the whole of frames signatures. Figure 4 illustrates our spatial-temporal signature generation process.

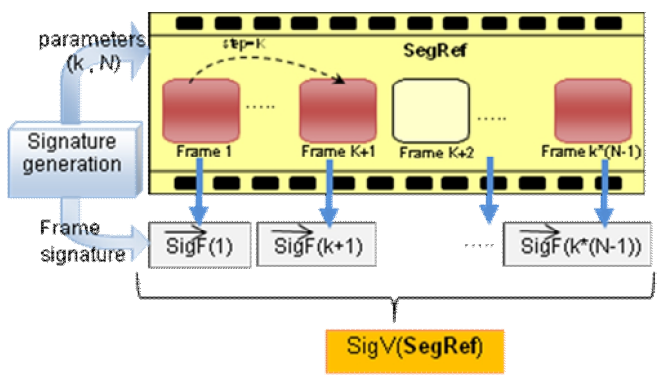

**Figure 4.** Process of spatial-temporal video signature generation.

### 3.1.4 Frame level signature

Several methods were proposed to create an identified image representation, especially these designed for CBIR applications. Most common approaches are based on low level features (color, intensity …) as detailed in the previous section. In our approach and in order to ensure the robustness of signature, we opted to use two descriptors: colorimetric feature and a *POI* descriptor.

#### a) CCV descriptor

For the colorimetric descriptor, histograms are used to represent images in many multimedia applications. Their advantages are insensitivity to small changes. However, color histograms lack spatial information, so images with very different appearances can have similar histograms. Hence, we use a histogram-based method for representing images, that incorporates spatial information. Each pixel is classified in a given color bucket as coherent or incoherent, based on whether or not it is part of a large similarly-homogeneous region. A color coherence vector (*CCV*) stores the number of coherent versus incoherent pixels for each color. *CCV* is a more sophisticated form of histogram refinement, in which buckets are partitioned based on spatial coherence. By separating coherent pixels from incoherent ones, *CCV* provides finer distinctions than classic histograms. The *CCV* computing process is composed essentially of three steps:

- *Image preprocessing*

This first step smoothes the image by applying a medium filter to the neighboring pixels. The major aim of this preprocessing is to eliminate small variations between adjacent pixels. Then, we precede to discrete the color space, to obtain only $N_{color}$ distinct colors in the image for the following two reasons: First decreasing the luminance variation effects, and secondly reducing the size of image signature.

- *Image segmentation*

To classify pixels within a given color bucket as coherent or incoherent, we proceed in the second step to region segmentation.

In order to determine the pixel groups, the image is segmented into disjoint and homogeneous regions. A region is defined as a connected set of pixels for which uniformity (homogeneity) condition is satisfied. Referring to segmentation method category, a uniform region can be obtained by two different ways: It can be derived from growing from a seed block/pixel by joining others pixels or obtained by splitting a large region which is not uniform.

Several image segmentation advanced techniques have been proposed and classified in different categories (region growing, split and merge…). We use the *Statistical Region Merging* (*SRM*) [13] algorithm that belongs to the family of region growing techniques with statistical test for region fusion. The advantages of this method are simplicity and performance without the use of color space transformations. In addition, we opted also for the *SRM* method as it gives, for each segmented region, the list of pixels belonging to it and the related mean color which facilitates afterward the computing of *CCV*'s bins. Note that in our work we apply the *SRM* method on grayscale quantified images. Image is segmented according to color buckets. This effectively segments the image based on the discretized color space.

The *SRM* method is based on two major components: a merging predicate (2) and the order followed in testing this predicate. The merging predicate is defined as:

$$P(R, R') = \begin{cases} true, & if \ \left| \overline{R'} - \overline{R} \right| \le b(R') + b(R) \\ false, & otherwise \end{cases} \qquad (2)$$

with: $b(R) = g \sqrt{\dfrac{1}{2Q|R|} \left( \ln \dfrac{R_{|R|}}{d} \right)}$

where $R$ and $R'$ represent the two regions being tested, $\overline{R}$ denotes the color average in region $R$ and $R_{|R|}$ is the set of regions with $p$ pixels. The order of region merging follows a criterion $f$, which implies that when any test between two parts within a true region is performed. $g$ and $Q$ are global (random) variables are used to determine the merging predicate threshold.

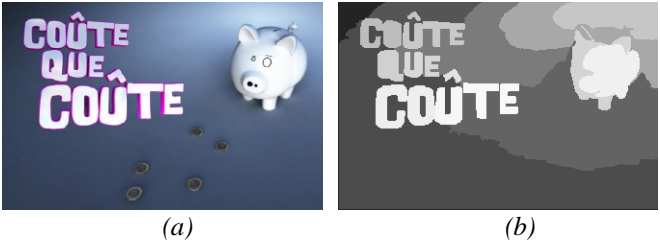

*(a)*           *(b)*

**Figure 5.** Image segmentation using SRM: (a) input color image, (b) segmented grayscale image.

The *SRM*'s algorithm is described as follow:

- First of all, an appropriate data structure was created to stack the pairs of adjacent pixels based on 4-connectivity.

  Let *p* and *p'* be two pixels of an image *I*, we denote by *R(p)* the region including the pixel *p*.

- Sort all pairs of pixels (using bucket sort) in ascending order according to a function *f (3)* which is a simplified version to calculate adjacent pixels gradients.

- After sorting, the test the predicate merge *P (R (p), R (p '))* only once for pixels for which *R (p) = R (p')* (*i.e.* belong to the same region) if the predicate is satisfied, the two regions *R (p)* and *R (p ')* will be merged. The merging phase is realized using the union-find algorithm (a simple and efficient algorithm to classify a set of elements into disjoint classes).

$$f(p, p') = |p'-p| \qquad (3)$$

- ***Pixels classification***

Following the image segmentation, each pixel belongs to exactly one region. Pixels are classified as coherent or incoherent depending on the size of their regions. A pixel is coherent if the size of region which it belongs exceeds a threshold (proportional to image size); otherwise, the pixel is considered as incoherent. For a given discretized color, some pixels will be coherent and some others will be incoherent. We nominate by $\alpha_j$ the number of coherent pixels of the *j'th* discretized color and by $\beta_j$ the number of incoherent pixels. The total number of pixels for color *j* is $\alpha_j + \beta_j$.

As a result, we obtain for each color *j* the pair $(\alpha_j, \beta_j)$ called the *j'th* color's coherence pair. The color coherence vector for the image consists of $((\alpha_1, \beta_1),....,(\alpha_n, \beta_n))$.

So, the set of bins presents the *CCV* frame signature (4):

$$SigF_{CCV} = \left\{ \overrightarrow{CCV_i}, i \in [1, N_{Color}] \right\} \qquad (4)$$

where $\overrightarrow{CCV_i} = (a_i, b_i)$

*b) POI descriptor*

We opted to use the *POI* (Point Of Interest) descriptor for the next reasons:

In image processing, *POI* is a point in an image that has special properties which make it stand out in comparison to its adjacent points. Referring to their definition, *POIs* are located where the photometric information is the most important within an image. These points are characterized by robustness to luminosity variation, blur effect and geometric transformations. Indeed, using a *POI* descriptor, we can provide a rich and compact signature at once.

Furthermore, studies of comparison between different *POI* detectors under variant conditions like illumination variation proved that most repeatable results are obtained for the Harris detector [6]. It is among the most stable and broadly used *POI* detectors due to its robustness to noise and lighting variations. The Harris detector relies on a basic principle: at a corner, the intensity will change greatly in different directions. Based on this observation, the central idea of this detector is to use the autocorrelation function, *i.e.* the second moment matrix, to determine locations where the image intensity changes in two directions.

Thus, the *POI*'s frame signature is expressed by a *POIs* vector (5):

$$SigF_{POI} = \left\{ \overrightarrow{POI_k}, k \in [1, N_{POI}] \right\} \qquad (5)$$

Where

$$\overrightarrow{POI_k} = \begin{pmatrix} x_k \\ y_k \\ r(x_k, y_k) \end{pmatrix},$$

*r(x,y)* is the Harris metric value at *(x,y)* pixel.

### 3.2 TV programs type identification

In this section, we describe the video programs identification process in TV streams. The basic idea of this process consists in comparing the visual similarity of the video signal at instant *k* to the catalogue reference elements. Our video type identification relies on a central assumption: two video $V_1$ and $V_2$ are visually similar if and only if their signatures are similar. The identification process is described as follows:

To identify the video type at the current frame *k*, for each video signature ***Vsig(t)*** selected from the catalogue, we load its $N_{FSig}$ frame signatures ***SigF*[1.. $N_{FSig}$]** where ***SigF*** denotes both ***SigF_{POI}*** and ***SigF_{CCV}***. For each elementary signature ***SigF*** (*j*: from *1* to $N_{FSig}$), we measure the similarity of this signature and the frame *k* signature.

#### 3.2.1 Signature descriptors similarity

The program TV type identification relies essentially on signature similarity measurement respectively between ***SigF_{CCV}*** descriptors and ***SigF_{POI}*** descriptors in reference catalogue and TV stream signal.

For ***SigF_{CCV}***, we calculate a normalized average of *CCV* bins similarities defined in frame signature ***SimSigF_{CCV}*** (6).

$$SimSigf_{CCV}(k,j) = \frac{\sum \left( \overrightarrow{CCV_k}, \overrightarrow{CCV_j} \right) - \left( \sum_{i=1}^{N_{color}} |a_k(i) - a_j(i)| + |b_k(i) - b_j(i)| \right)}{\sum \left( \overrightarrow{CCV_k}, \overrightarrow{CCV_j} \right)} \qquad (6)$$

with :
$$\sum (\overrightarrow{CCV}_k, \overrightarrow{CCV}_j) = \left( \sum_{i=1}^{N_{cca}} a_k(i), b_j(i), a_j(i), b_k(i) \right)$$

As for $SigF_{POI}$ signature similarity measurement, we test the Euclidean distance of all *POIs* values defined in this signature. Two $SigF_{POI}$ signatures are similar if and only if the majority of *POIs* (a percentage number of $N_{POI}$) descriptors are similar (7).

$$SimSigf_{POI}(k,j) = \frac{\sum_{k=1}^{N_{POI}} Sim(POI_k, POI_j)}{N_{POI}} \qquad (7)$$

with:

$$Sim(\overrightarrow{POI_k}, \overrightarrow{POI_j}) = \begin{cases} 1, & if \; Dist_{eucd} < threDist \, \& Dist_{harris} < threHarris \\ 0, & otherwise \end{cases}$$

$$Dist_{eucd}(\overrightarrow{POI}_k, \overrightarrow{POI}_j) = \sqrt{(x_k - x_j)^2 + (y_k - y_j)^2}$$

$$Dist_{Harris}(\overrightarrow{POI}_k, \overrightarrow{POI}_j) = \left| r(x_k, y_k) - r(x_j, y_j) \right|$$

### 3.2.2 Descriptors similarity combination

Since a frame has a composed-signature ($Sigf_{CCV}$ and $Sigf_{POI}$), to detect the video type after computing the similarity of these signatures, we combine their similarities to obtain a single decision value: $Sim_{Video}$(9). This combination aims at increasing the identification rates.

The combination is done using the average rule, following normalizing and pondering coefficients of two signature descriptors. Hence, the frame signature similarity is defined as (8):

$$simSigf(k,t) = \frac{w_1 \times simSigf_{CCV}(k,t) + w_2 \times simSigf_{POI}(k,t)}{w_1 + w_2} \qquad (8)$$

$$sim_{Video}(k,t) = \begin{cases} 1, if \; simsigf > th_{Sig} \\ 0, ohterwinse \end{cases} \qquad (9)$$

with:

$$th_{Sig} = \frac{w_1 th_{POI} + w_2 th_{CCV}}{w_1 + w_2}$$

According to our experimental study, we concluded that a discriminative identification of a descriptor *d* differ from a channel to another. This could be explained by the fact that the graphic charter of channel used during the generics production stage exploits graphics compositions rich of a particular descriptor (Figure 3) than another. Therefore, the weighted coefficient $w_d$ of descriptor *d* must be relative to its efficiency to identify individual types of programs to indicate the importance of *d* in the combination with other(s) descriptor(s) (8). In other words, the identification rate of *d* is as important as its weight of its coefficient. That is, this weight must be proportional to identification rate (12).

The relevance of descriptor *d* is evaluated by its capacity to identify the maximum number of correct identifications with a minimum number of false identifications *i.e.* *d* is significant if both recall and precision having high values at once. So that, $w_d$ is a combination formula of these two metrics.

$$R = \frac{CI_d}{CI_d + MI_d} \qquad (10)$$

$$P = \frac{CI_d}{CI_d + FI_d} \qquad (11)$$

$$w_d = F1 = \frac{2 \times R \times P}{R + P} \qquad (12)$$

With:
$CI_d$: number of programs identified correctly using descriptor *d*
$MI_d$: number of missed identifications using descriptor *d*
$FI_d$: number of false identifications using descriptor *d*

In order to define these weights, we conducted a training phase to choose the optimum weight value for the descriptor *d* of each TV program/channel. Table 1 summarizes the weight values for various TV channels.

**Table 1**: $w_1$ and $w_2$ values for each channel.

| Descriptor Channel | Recall (*R*) | | Precision (*P*) | | $w_d$ (*F1*) | |
|---|---|---|---|---|---|---|
| | $d_1$ | $d_2$ | $d_1$ | $d_2$ | $w_1$ | $w_2$ |
| M6 | 0,8 | 0,92 | 1 | 0,92 | 0,89 | 0,92 |
| RTV | 0,95 | 0,66 | 0,75 | 0,87 | 0,84 | 0,75 |
| LCI | 0,99 | 0,69 | 0,98 | 0,9 | 0,98 | 0,78 |
| itele | 0,83 | 0,53 | 0,67 | 0,75 | 0,74 | 0,62 |
| Abmoteurs | 0,66 | 0,74 | 1 | 0,67 | 0,79 | 0,70 |
| France 24 | 1 | 0,5 | 0,84 | 0,71 | 0,91 | 0,59 |

A segment localized at frame *j* in a video stream was identified of type *t* only if the frames signatures ($N_{sigF}$) of *Vsig(t)* are similar to their homologous in the stream as detailed in our similarity measurement metrics (13).

$$VideoSegment(j) = \begin{cases} videoType(t), & if \; \sum_{i=1}^{N_{sigF}} Sim_{Video}(k,t) = N_{sigF} \\ undefined, & otherwise \end{cases} \qquad (13)$$

## 4. Experimental Results

### 4.1 AViTyp: Automatic Video Type identification tool

To implement the proposed approach and in order to evaluate its efficacy, we have developed a system called *AViTyp* (Figure 6). This system offers two main features: signature creation for references catalogue items, and programs identification in files from TV channels. In addition, *AViTyp* provides an ergonomic interface to adjust identification process: defining the weights and appropriate thresholds for the current channel.

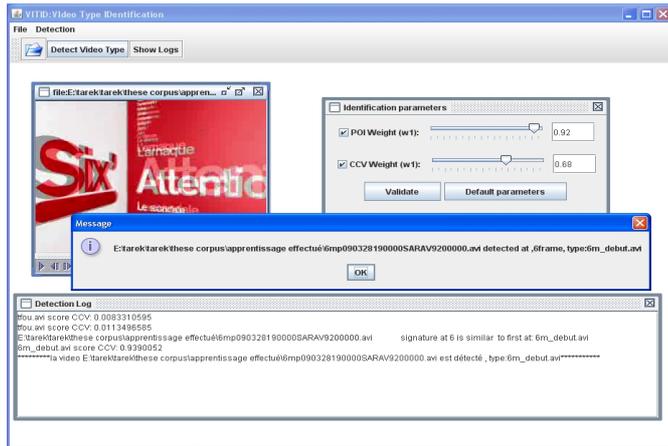

**Figure 6.** User interface of the *AViTyp* tool.

### 4.2 Evaluation of the video type identification performance

To evaluate experimentally our video type identification approach, we used a large and varied corpus composed of a set of video files which are long streams from 6 different TV channels. Various programs and inter-programs types (news, sport, varieties, documentaries, pubs…) are contained in these streams.

To evaluate the performance of the proposed approach, we used the recall (10) and precision (11) metrics. Table 2 presents the experimental results values regrouped by TV channel.

**Table 2**: Experimental results grouped by channel.

| TV channel | recall | Precision |
|---|---|---|
| M6 | 93,2 | 100 |
| RTV | 81,6 | 100 |
| LCI | 85,22 | 85,71 |
| Itele | 74,2 | 83,33 |
| Abmoteurs | 75 | 100 |
| France 24 | 94,7 | 87,5 |
| **All channels (average)** | **83,98** | **92,75** |

In this experimentation, despite the good precision value, we conclude that the recall was rather satisfactory (83,98%) and needs to be improved. The degradation of the recall is due essentially to some missed identifications. The main reason behind the missed cases is due essentially to the quality of the broadcast streams, like a blur signal.

## 5. Conclusion and future work

We have proposed in this paper a grammar-based approach for video program identification in TV streams. The approach is composed by two steps: (i) creation of references catalogue and (ii) identification of programs TV in channels streams. It compares the visual similarity between TV stream signal and video signatures stored within as grammar descriptors in a reference catalogue. This catalogue is composed of a set of visual "jingles" that characterize the starting of TV programs associated, with their grammars expressed as spatial-temporal video signatures.

In order to improve the video type identification quality, we focus our future work on the integration of other descriptors such as the form or the texture features in video grammar which can characterize the visual jingles since the used features are not always discriminative.

## Authors Profile

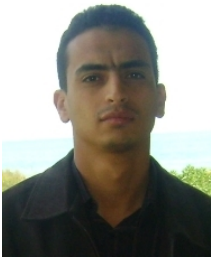

**Tarek ZLITNI** received the M.S degree in information systems and new technologies in 2007 from the University of Sfax, TUNISIA, where he is pursuing the Ph.D. degree in computer science.
His research interests focus on video and image processing and analysis, multimedia indexing, and content-based video segmentation and structuring.

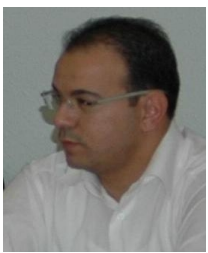

**Walid MAHDI** received a Ph.D. in Computer and Information Science from *Ecole Centrale de Lyon, France* in 2001. He is currently Assistant Professor at Higher Institute of Computer Science and Multimedia, at the University of Sfax, TUNISIA. His research is about image and video processing.